\newcommand{\La}{La$_{2}$CuO$_{4}$}
\newcommand{\LaSr}{La$_{1.86}$Sr$_{0.14}$CuO$_{4}$}
\newcommand{\LaSrx}{La$_{\rm 2-x}$Sr$_{\rm x}$CuO$_{\rm 4}$}
\newcommand{\chiQw}{$\chi^{\prime\prime}({\bf Q},\omega)$}
\newcommand{\SQ}{$S({\bf Q})$}
\newcommand{\SQw}{$S({\bf Q},\omega)$}
\newcommand{\chiQ}{$\chi^{\prime}({\bf Q})$}
\newcommand{\chiw}{$\chi^{\prime\prime}({\omega})$}

\documentstyle[preprint,tighten,aps]{revtex}
\begin{document}
\draft

\title{A Comparison of the High-Frequency Magnetic Fluctuations in Insulating and Superconducting \LaSrx}

\author{S.\ M.\ Hayden$^{(1)}$, G.\ Aeppli$^{(2,3)}$, H.\ A.\ Mook$^{(4)}$ 
, T. G.  Perring$^{(5)}$, T.\ E.\ Mason$^{(6)}$, S-W. Cheong$^{(2)}$ and Z.\ 
Fisk$^{(7)}$} \address{$^{(1)}$H.\ H.\ Wills Physics Laboratory, University
of Bristol, Tyndall Avenue, Bristol BS8\ 1TL, United Kingdom\\
$^{(2)}$AT\&T Bell Laboratories, Murray Hill, New Jersey 07974\\
$^{(3)}$Ris\o\ National Laboratory, 4000 Roskilde, Denmark\\
$^{(4)}$Oak Ridge National Laboratory, Oak Ridge, Tennessee 37831\\
$^{(5)}$ISIS Facility, Rutherford Appleton Laboratory, Chilton, 
Didcot,OX11\ 0QX, United Kingdom\\ 
$^{(6)}$Department of Physics, University of
Toronto, Toronto, M5S\ 1A7, Canada\\ 
$^{(7)}$Department of Physics, Florida
State University, Tallahassee, Florida 32306}

\date{21 November, 1995}
\maketitle

\begin{abstract}
Inelastic neutron scattering performed at a spallation source is used to
make absolute measurements of the dynamic susceptibility of insulating \La\ and superconducting \LaSr\ over
the energy range $15 \leq \hbar\omega \leq 350$ meV.   
The effect of Sr doping on the magnetic excitations is to cause a large broadening in 
wavevector and a substantial  change in the spectrum of the local spin fluctuations.  
Comparison of the two compositions reveals a new energy scale $\hbar\Gamma = 22\pm5$\ meV in \LaSr.
\end{abstract}

\narrowtext

It is well known that the parent compounds of high temperature 
superconductors are antiferromagnets displaying strong spin 
fluctuations\cite{Shirane87}.  
The one-magnon excitations of one such parent compound, the 
spin-$\case{1}{2}$ square-lattice antiferromagnet \La, have been 
characterized throughout the Brillouin zone\cite{Aeppli89,Hayden91,Itoh94}.
In contrast, observations 
of the magnetic excitations for superconducting 
compositions\cite{Cheong91,Matsuda94} have been 
limited to relatively low frequencies.  Indeed, the measuring frequencies have generally been below pairing energies ($\approx 10$\ meV for optimally doped \LaSrx) and very much less than the underlying magnetic coupling strengths and spin-flip energies $2J \approx 320$\ meV of the parent insulating antiferromagnets.  Thus,  the extent to which doping affects the underlying antiferromagnetism is unknown for energies beyond $J/10$.  Exploring this energy range is essential because it is the short-wavelength and high-frequency spin physics which 
may ultimately be responsible for many of the unusual properties of the cuprates.  One might ask whether this physics is the same in the superconductors and insulating parents. To begin to answer such questions, we have used neutron scattering to measure the magnetic response of \LaSrx\ with $x=0.14$ throughout the Brillouin zone and for energy transfers 
between $J/10$ and $2.5J$.  
Neutron scattering is unique in comparison to other probes\cite{Uchida91,Imai93,Raman} of the magnetic response both because of its unrivalled access to excitations with arbitrary energy and wavevector as well as a magnetic cross-section which is simple and well-known. Our key result is that the doping affects the magnetic fluctuations
at all frequencies up to $2J \approx 320$\ meV.  In particular, spectral weight is removed from the magnetic Bragg peak and the high-frequency spin waves of the pure \La\ parent to create a magnetic fluctuation spectrum with a distinct maximum at an energy of order 20\ meV.     

Experiments were performed on the MARI spectrometer at the ISIS spallation 
neutron source of the Rutherford-Appleton Laboratory.  MARI is a 
direct geometry chopper spectrometer.  The use of this type of spectrometer 
for measuring high-frequency spin fluctuations in single crystals has been 
described elsewhere\cite{Hayden91,Itoh94}.  The samples in the present 
investigation were assemblies of single crystals with total masses 48.6g 
and 24g for \La\ and \LaSr\ respectively, and were prepared as for previous 
experiments\cite{Hayden91,Cheong91}.  The $x=0.14$ crystals were all bulk 
superconductors with $T_{\text{c}}=35 \text{ K}$.  Following previous 
practice\cite{Shirane87,Aeppli89,Hayden91,Matsuda94}, we use the orthorhombic nomenclature to label 
reciprocal space so that the basal planes are parallel to the (010) plane.  
We do not distinguish between $a$ and $c$.  In this notation, 
antiferromagnetic order in \La\ occurs at ${\bf Q}=(1,0,0)$.  For all data 
reported here, samples were aligned with the (001) plane coincident with 
the principal scattering plane of the spectrometer.  The detectors were 
300mm-long 25mm-diameter tubes arranged with their axes perpendicular to 
the spectrometer plane.  Scattering angles $2\theta$ were in the range 3.43 
to 15.0 deg.  Following common practice, absolute unit conversions were performed using a vanadium standard\cite{Itoh94,Windsor81}. 

Fig.\ \ref{fig1} shows data collected as a function of ${\bf Q}$ 
along the $(1,0,0)$ direction for various energy transfers 
$\hbar\omega$ with $E_{i}=$300\ meV.  For pure \La, this implies that we intercept the magnon dispersion surface along horizontal trajectories of the type shown in 
Fig.\ \ref{fig1}(a).  The magnetic neutron scattering cross section per formula unit is given by\cite{Lovesey84},
\begin{eqnarray}
\label{eq1}
\frac{d^2\sigma}{d\Omega \: dE} &=& (\gamma r_{\text{e}})^2 \frac{k_f}{k_i}
\left| F(Q)\right|^2  \nonumber \\ 
&& \times \frac{1}{\hbar} \sum_{\alpha\beta} (\delta_{\alpha\beta}-\hat{Q}_{\alpha}\hat{Q}_{\beta})
S^{\alpha\beta} ({\bf Q},\omega)\;,
\end{eqnarray}
where $k_{i}$ and $k_{f}$ are the initial and final neutron wavenumbers respectively, and $S^{\alpha\beta} ({\bf Q},\omega)$, the Fourier transform of the spin-spin correlation function, is  
\begin{eqnarray}
\label{eq2}
S^{\alpha\beta} ({\bf Q},\omega) &=& \frac{1}{2\pi} \int dt \, e^{i \omega t} 
\frac{1}{N}\sum_{n,m} e^{i {\bf Q} \cdot ({\bf r}_{\scriptstyle m} - {\bf r}_{\scriptstyle n})} \nonumber \\
&& \times  \left< S^{\alpha}_{n}(t) \, S^{\beta}_{m}(0) \right>\;.
\end{eqnarray}
For an antiferromagnet, such as \La, we average over magnetic domains yielding 
$S({\bf Q},\omega)=\frac{1}{3} \left[S^{xx}({\bf Q},\omega)+S^{yy}({\bf Q},\omega)+S^{zz}({\bf Q},\omega)\right]$.  Because of the large variation of the $k_f/k_i$ Fermi factor and Cu$^{2+}$ magnetic form factor $F(Q)$ \cite{Shirane87,Brown92} over the data in Fig.\ \ref{fig1}, we have transformed our results using Eq.\ 1 to yield \SQw.
Our observations are consistent with previous 
lower-statistics data collected on a similar spectrometer\cite{Hayden91}.  
As Fig.\ \ref{fig1}(b)-(f) illustrates, a spin-wave peak is observed near 
$h=1$.  The peak becomes broader at higher energies due to the spin-wave dispersion.  Twin peaks due to counter-propagating spin-wave 
branches are not seen due to the poor resolution in the out-of-plane (001) 
direction.  The increased scattering at larger $h$ is due to phonons.

Fig.\ \ref{fig1}(h)-(l) shows analogous data for \LaSr\ collected under the same 
spectrometer conditions as Fig.\ \ref{fig1}(b)-(f) {\it and in the same units}.  
At all frequencies, the scattering for the metal, \LaSr, is substantially broader than for the 
insulator, \La.  
This is not surprising given that, at low $\omega$, the metal displays 
incommensurate magnetic peaks, with 
$h=0.88 \text{ and } 1.12$\cite{Cheong91,Matsuda94} when 
projected onto the (1,0,0) direction.  These peaks broaden rapidly with 
increasing $\hbar\omega$.  In contrast to what occurs for the insulator, the broadening in 
${\bf Q}$ is not obviously dependent on $\hbar\omega$.  

At the highest energy transfers in Fig.\ 1, it is clear that the signal is vanishing more rapidly for the superconductor than for the insulator.
We therefore increased the sensitivity of the spectrometer, while coarsening the resolution, by raising $E_{i}$ to 600\ meV.  
Fig.\ \ref{fig2}(a)--(c) shows the resulting cross-section, corrected only for the Fermi factor in Eq.\ 1, in the form of constant-$\omega$ scans (horizontal trajectories in Fig.\ 1(g)).  
For $\hbar\omega=175\pm25$\ meV (frame(c)), we find that a magnetic peak remains at (1,0,0).
A peak is no longer apparent for $\hbar\omega=237.5 \pm 37.5$\ meV, but there is still finite scattering.  
Finally, for $\hbar\omega=325\pm50$\ meV, the scattering is indistinguishable from zero.
Fig.\ 2(d) and (e) show the $E_{i}=\text{ 600 meV}$ data transformed to give \SQw\ and plotted  in the form of constant-${\bf Q}$ scans centered on the magnetic zone center (1,0,0) and zone boundary (1.5,0,0) (vertical trajectories in Fig.\ \ref{fig1}(g)).  
The important conclusion from both the constant-$\hbar\omega$ and the constant-${\bf Q}$ scans is that there is appreciable magnetic scattering up to but not beyond a cut-off of approximately 280\ meV.   Furthermore, there is a weak peak near 240\ 
meV for the zone-boundary scan Fig.\ \ref{fig2}(d).   The observed scattering is approximately three times weaker (dotted line) than expected from \La\ under the same experimental conditions.  It has also been shifted to a frequency 20\% lower than in the insulating 
parent\cite{Aeppli89,Hayden91,Itoh94}.  

So far, we have concentrated on the detailed $\omega$- and ${\bf Q}$-dependence of the observed response.  
Also of interest are aggregate quantities such as (i) the equal-time spin correlation function $S({\bf Q})=\int_{-\infty}^{\infty} d\omega \: S({\bf Q},\omega)$; (ii) 
the zero-frequency susceptibility $\chi^{\prime}({\bf Q})=$ $(1/\pi) 
\int_{-\infty}^{\infty} d\omega \: \chi^{\prime\prime}({\bf 
Q},\omega)/\omega$; and (iii) the local susceptibility $ 
\chi^{\prime\prime}(\omega) =$$\int \chi^{\prime\prime}({\bf Q}, \omega) \; 
d^3Q / \int d^3Q$.   
We have therefore computed (i)-(iii) directly from our data, exploiting the 
fact that for 2D systems, such as \LaSrx, the experimental geometry chosen 
provides essentially complete images of $\chi^{\prime\prime}({\bf Q}_{\text{2D}},\omega)$\cite{Hayden91}.  
Our estimates are for the energy range 15-150\ meV, previous studies have been limited to energies below 30\ meV\cite{Shirane87,Cheong91,Matsuda94}.   
For \La, \SQ\ consists of a peak centered at (1,0,0) (Fig.\ \ref{fig3}(a)) with
half-width-at-half-maximum (HWHM) $\kappa=0.16 \pm 0.01 \text{ \AA}^{-1}$.  
The fact that the peak in \SQ\ is considerably broader for \LaSr (Fig.\ \ref{fig3}(c)), examined using the same spectrometer configuration, implies that the experimental resolution, for which no corrections have been made, makes a small contribution to $\kappa$ for \LaSr.  
The peak width is $\kappa=0.27 \pm 0.04 \text{ \AA}^{-1}$, corresponding to a pair correlation length of $3.7 \pm 0.5 \text{ \AA}$ which is indistinguishable from the nearest-neighbor separation ($3.8 \text{\AA}$) between copper atoms.
In the case of \La\ (Figs.\ \ref{fig3}(a) and (b)), \chiQ\ ($\kappa=0.13 \pm 0.01 \text{ \AA}^{-1}$) is somewhat sharper than \SQ\ ($\kappa=0.16 \pm 0.01 \text{ \AA}^{-1}$), 
because the Kramers-Kronig transform gives relatively higher weight to low frequencies 
where the spin wave peak is narrower (see Fig.\ \ref{fig1}).  
On the other hand, for \LaSr, where the width of the peak centered on (1,0,0) is not obviously 
$\omega$-dependent, the shapes of $\chi^{\prime}({\bf Q})$ and \SQ\ are 
virtually identical.  In particular, a HWHM of $\kappa=0.26\pm0.05 \text{ \AA}^{-1}$ 
characterizes \chiQ.  The Brillouin-zone-averaged susceptibility $\left< 
\chi^{\prime}({\bf Q})\right>_{BZ}$ derived from our $\hbar\omega \geq 15$\ meV 
measurements is $2.4 \pm 0.5 \: \mu_{\text{B}}^{2} \text{ eV}^{-1} \text{f.u.}^{-1}$ 
or $1.9 \pm 0.4 \times 10^{-7} \: \text{e.m.u. g}^{-1}$, which is comparable to the 
measured bulk $(Q=0)$ susceptibility\cite{Takagi89} $\chi_{\text{bulk}} \approx 1 \times 
10^{-7} \text{ e.m.u. g}^{-1}$.

Fig.\ \ref{fig4} shows that the local susceptibility 
\chiw\ evolves as dramatically with doping as \SQ\ and \chiQ.  
While \chiw\ increases slightly
with $\omega$ for pure \La, it decreases with $\omega$ over the frequency
range probed in the present experiment on \LaSr.  If we also include data
from a reactor-based experiment\cite{Aeppli95} on \LaSr\ we find that 
$\chi^{\prime\prime}(\omega)$ is actually peaked near 
$\hbar\omega=\hbar\Gamma = 22 \pm 5 \text{ meV}$ and a Lorentzian  
$\chi^{\prime\prime}(\omega) \sim \Gamma\omega / (\Gamma^2+\omega^2)$
describes the data. Of course, for the pure
compound, there is a magnetic Bragg peak (with weight 0.36\ $  
\mu_{\text{B}}^{2} \text{ fu}^{-1}$) at $\hbar\omega=0$, implying an underlying 
relaxation rate $\Gamma=0$.  Thus, doping results in a transfer of spectral weight
($\chi^{\prime\prime}(\omega)$) from near $\omega=0$ to a peak centered 
at an intermediate frequency of order 20\ meV.  
More surprising is the concomitant suppression of the high frequency magnetic signal.

The simplest theory to which we can compare our \La\ data is that of linear spin
waves\cite{Lovesey84}.  In a two-dimensional antiferromagnet, the spin-wave velocity and overall amplitude show quantum renormalizations\cite{Anderson52,Igarashi92} with respect to their classical (large $S$) values which can be described by the factors $Z_{c} \approx 1.18$ and $Z_{\chi}$ respectively.  Fig.\ 1(b)-(f) shows fits to the data of linear spin-wave theory with an effective exchange interaction $J=153$\ meV\cite{Hayden91}, including a frequency-independent quantum renormalization $Z_{\chi}$ of the overall intensity.  
Our measured $Z_{\chi} = 0.39 \pm 0.1$ is in agreement with previous work\cite{Itoh94} and close to the prediction of a $1/S$ spin-wave expansion\cite{Igarashi92}, $Z_{\chi} = 0.51$. 

Spin-wave theory cannot be used to describe the data for \LaSr\ because 
the scattering is both too broadly peaked in ${\bf Q}$ and strongly decreasing
with $\hbar\omega$ (see Fig.\ 4).  Even so, spin-wave theory may be applicable
at very short wavelengths and high frequencies.  
We have therefore fitted a linear spin-wave model to the data in Fig.\ \ref{fig2} for $\hbar\omega > 100 \text{ meV}$.  
The solid lines represent the best fits for which 
$J=130 \pm 5 \text{ meV}$ and $Z_{\chi}=0.15\pm0.06$.  While the effective $J$ is 
not much less than that for pure \La, $Z_{\chi}$ is much reduced.
This can be seen from the comparison between the expected signal for \La, calculated for the present spectrometer resolution using the previously established cross-section\cite{Hayden91}, , and the observed signal for \LaSr\ shown in Fig.\ \ref{fig2}(d). 

To place the current results in the context of what is known about the spin dynamics of \LaSr, we recall that at low frequencies, $\hbar\omega \lesssim 15$\ meV, \chiQw\ shows peaks at incommensurate positions around (100)\cite{Cheong91,Matsuda94}.  
The peaks broaden considerably as $\hbar\omega$ is increased towards 15\ meV.  It should therefore come as no surprise that we do not observe incommensurate maxima in the current high-energy measurements, for which $\hbar\omega \geq 15$\ meV.  
Comparison of the present data and low-$\omega$ data\cite{Aeppli95}, both of which are on absolute scales, shows that the incommensurate features contribute considerably less integrated weight, to $S({\bf Q})$, than the broad peak around (1,0,0) displayed in Fig.\ 3(c).  Thus it appears that the magnetic response for \LaSr\ consists of a quasiparticle contribution \cite{Littlewood93}, giving rise to sharp incommensurate features, superposed on a broad background due to short-range antiferromagnetic correlations.

In summary, our measurements on \LaSr\ show that doping the Mott insulator \La\ dramatically modifies its magnetic excitations over a wide frequency range.  The principal changes are 
that the excitations are greatly broadened in wavevector and there is a substantial redistribution of spectral weight in frequency.  The corresponding antiferromagnetic correlation length, deduced from examining integrals of \chiQw\ over an unprecedently large frequency range, is indistinguishable from the separation between nearest neighbor Cu atoms. 
Finally, when combined with earlier data, the present results imply a new energy scale,
$22 \pm 5 \text{ meV}$, characterizing the spin fluctuations in superconducting \LaSr.
 
We are grateful to M. Lund for technical assistance, Ris\o\ National Laboratory for hospitality during the preparations for this experiment and to EPSRC, NSERC, CIAR and US-DOE  for financial support.


\begin{figure}
\caption{(a)-(f) Magnetic scattering from \La\ for $E_{i}=300$ meV and ${\bf k}_{i}
\parallel (0,1,0)$. The sample mass was 48.6g and the
counting time was 9\ h at 170\ $\mu$A proton current with Ta 
target. The broadening in $h$ at higher energies is due to the 
spin-wave dispersion.   Solid lines are fits to a resolution-corrected spin-wave
cross section.  
The scans shown in the figure are the raw data minus an estimate of the background, namely the intensity measured at the lowest accessible scattering angles $3.8 \leq 2\theta \leq 4.3$\ deg, where the scattering from the antiferromagnet is expected to be small. 
(g)-(l) Magnetic scattering in \LaSr. Incident energy and crystal orientation 
as in (a)-(f). The sample mass was  24g and the counting time was 43 h.
Note the scattering is broader in ${\bf Q}$ than that observed 
for \La.  Solid lines are a resolution-corrected momentum-broadened 
spin-wave cross section.}
\label{fig1}
\end{figure}

\begin{figure}
\caption{Magnetic scattering from \LaSr\ for $E_{i}=600$ meV and ${\bf k}_{i}
\parallel (0,1,0)$.  Counting time was 81 h. (a)--(c) are constant-energy scans across the ridge of magnetic scattering illustrated in Fig.\ 1(g).  (d) and (e) are constant-${\bf Q}$ scans centered on the magnetic zone center (ie ${\bf Q}=(1,0,0)$) and zone boundary (${\bf Q}$=(1.5,0,0)) respectively. 
Detectors with the same average $2\theta$, where the antiferromagnetic
structure factor is small, have been used as a background. Solid lines in all frames are for the spin-wave model described in the text.  Dotted line is scattering expected for \La\ based on previous 
measurements\protect\cite{Hayden91}.  }
\label{fig2}
\end{figure}

\begin{figure}
\caption{Local susceptibility $\chi^{\prime\prime}(\omega)$ determined from
integration over wavevector of the observed scattering for (a) \La\ and (b) \LaSr.  
Open circles, data from reactor-based experiments\protect\cite{Aeppli95}; closed circles are the present work.} 
\label{fig3}
\end{figure}

\begin{figure}
\caption{ The Fourier transform of the equal-time correlation function 
\SQ\ and the real part of the wavevector-dependent susceptibility
\chiQ\ for \La\ ((a) and (b)) and \LaSr\ ((c) and (d)) respectively.  These
quantities have been determined from our measurements in the
energy range $15 \leq \hbar\omega \leq150$ meV.}
\label{fig4}
\end{figure}


\begin{references}
\bibitem{Shirane87}D. Vaknin {\it et al.}, Phys. Rev. Lett. {\bf 58}, 2802 (1987); G. Shirane {\it et al.}, {\it ibid.} {\bf 59}, 1613 (1987); J. M. Tranquada {\it et al.}, {\it ibid.} {\bf 60} 156 (1988); J. Rossat-Mignod {\it et al.}, J. Phys. (Paris) {\bf 49}, C8-2119 (1988).     
\bibitem{Aeppli89}G. Aeppli {\it et al.}, Phys. Rev. Lett. {\bf 62}, 2052 (1989).  
\bibitem{Hayden91}S. M. Hayden {\it et al.}, Phys.  Rev.  Lett.  {\bf 67}, 3622 (1991).
\bibitem{Itoh94}S. Itoh {\it et al.}, J. Phys. Soc. Japan {\bf 63}, 4542 (1994). 
\bibitem{Cheong91}S.-W. Cheong {\it et al.}, Phys. Rev. Lett. {\bf 67}, 1791 (1991); T. E. Mason, G.\ Aeppli and H.\ A.\ Mook, {\it ibid.} {\bf 68} 1414 (1992).   
\bibitem{Matsuda94}T. R. Thurston {\it et al.}, Phys. Rev. B {\bf 46}, 9128 (1992); M. Matsuda {\it et al.}, {\it ibid.} {\bf 49}, 6958 (1994).
\bibitem{Uchida91}S. Uchida {\it et al.}, Phys. Rev. B {\bf 43}, 7942 (1991).
\bibitem{Imai93}T. Imai {\it et al.}, Phys. Rev. Lett. {\bf 70}, 1002 (1993); R. E. Walstedt {\it et al.}, {\it ibid.} {\bf 72}, 3610 (1994).
\bibitem{Raman}K. B. Lyons {\it et al.}, Phys. Rev. Lett. {\bf 60}, 1317 (1988); E. T. Heyen {\it et al.}, Phys. Rev. B {\bf 43}, 12958 (1991); G. Blumberg {\it et al.}, {\it ibid.} {\bf 49}, 13295 (1994).
\bibitem{Windsor81}C. G. Windsor, {\it Pulsed Neutron Scattering} (Taylor and Francis, London, 1981)
\bibitem{Lovesey84}S. W. Lovesey, {\it Theory of Neutron Scattering from Condensed Matter} Vol. 2 (Oxford, 1984).
\bibitem{Brown92}T. Freltoft {\it et al.}, Phys. Rev. B {\bf 37}, 137 (1988); P. J. Brown, in {\it International Tables for Crystallography} Vol. C, edited by A. J. C. Wilson (Kluwer, Dordrecht, 1992).
\bibitem{Takagi89}D. C. Johnston {\it et al.}, Physica C {\bf 153-155}, 572 (1988); H. Takagi {\it et al.}, Phys. Rev. B {\bf 40}, 2254 (1989). 
\bibitem{Aeppli95}G. Aeppli {\it et al.}, submitted to Phys. Rev. B (1994).
\bibitem{Anderson52}P. W. Anderson, Phys. Rev. {\bf 86}, 694 (1952); R. Kubo,
{\it ibid.} {\bf 87}, 568 (1952); T. Oguchi, {\it ibid.} {\bf 117}, 117 (1960).
\bibitem{Igarashi92}J. Igarashi, Phys. Rev. B {\bf 46}, 10763 (1992) and J. Physics: Cond. Matt. {\bf 4} 10265 (1992).
\bibitem{Littlewood93}See e.g. N. Bulut and D. J. Scalapino, Phys. Rev. B {\bf 45},
2371 (1992); Q.\ Si, Y.\ Zha and K.\ Levin, J. Appl. Phys. {\bf 76}, 6935 
(1994); P. Benard, L. Chen and A.-M. S. Tremblay, Phys. Rev. B {\bf 47}, 589 (1993); P. B. Littlewood {\it et al.}, {\it ibid.} {\bf 48}, 487 (1993).  
\end{references}
\end{document}